\documentstyle[twocolumn,aps,prl,epsfig]{revtex}

\newcommand{\srbsix}{SrB$_{6}$}
\newcommand{\cabsix}{CaB$_{6}$}

\newcommand{\calahalf}{Ca$_{0.995}$La$_{0.005}$B$_{6}$}

\begin{document}

\twocolumn[\hsize\textwidth\columnwidth\hsize\csname@twocolumnfalse\endcsname

\author{Donavan Hall$^{1}$, D. P. Young$^{1,2}$, Z. Fisk$^{1}$, 
 T. P. Murphy$^{1}$, E. C. Palm$^{1}$, and A. Teklu$^{2,3}$, R. G. Goodrich$^{2}$}
\address  {$^{1}${\small \ National High Magnetic Field
Laboratory, Florida State University, Tallahassee, FL 32310.}}
\address{ $^{2}${\small Department of Physics and Astronomy,
Louisiana State University, Baton Rouge, LA 70803.}}
\address{ $^{3}${\small National Center for Physical Acoustics,
University of Mississippi, University, MS 38677.}}
\title{Fermi Surface Measurements on the Low Carrier Density
Ferromagnet
    Ca$_{1-x}$La$_{x}$B$_{6}$ and SrB$_{6}$}
\date{April 5, 2001}
\
\maketitle
\begin{abstract}
Recently it has been discovered that weak ferromagnetism of a dilute
3D electron gas develops on the energy scale of the Fermi
temperature in some of the hexaborides; that is, the Curie temperature
approximately equals the Fermi temperature.  We report the results of
de Haas-van Alphen experiments on two concentrations of La-doped
CaB$_{6}$ as well as Ca-deficient Ca$_{1-\delta}$B$_{6}$ and
Sr-deficient Sr$_{1-\delta}$B$_{6}$.  The results show that a Fermi
surface exists in each case and that there are significant
electron-electron interactions in the low density electron gas.
\end{abstract}
\\
\maketitle
]


The cubic hexaborides of the alkaline and rare earth elements have long
attracted interest for their wide variety of physical properties in spite of
the simple crystallographic structure. These physical properties include
very low work functions leading to the use of LaB$_{6}$ as a thermionic
emitter, dense Kondo behavior and electric quadrupole ordering in CeB$_{6}$,
Kondo insulating properties in SmB$_{6}$ and interesting low carrier density
ferromagnetism in the local moment system EuB$_{6}$. A new aspect to the
rich spectrum of properties found in the hexaborides, high temperature weak
ferromagnetism at low carrier concentration (HTFLCC) with no atomic
localized moments was added about two years ago\cite{YoungNature}.

The host materials for HTFLCC are the divalent alkaline earth hexaborides 
CaB$_{6}$, SrB$_{6}$ and BaB$_{6}$. The crystal structure of these materials
can be thought of as a simple cubic CsCl-type arrangement of B$_{6}$%
-octaheda and metal ions. The early electronic structure cluster
calculations of Longuet-Higgins and Roberts\cite{Longuet} found that the
linked B$_{6}$-network required 20 electrons for a closed shell electronic
configuration, indicating that the alkaline earth hexaborides would be
semiconductors. More recent band structure calculations\cite{Massida} show
that divalent hexaborides should be semimetals, with a small direct overlap
of a primarily boron derived valence band and with a primarily alkaline
earth derived conduction band at the X-point in the Brillouin zone. A study
of the low temperature properties of single crystals of SrB$_{6}$ did find
approximately 0.001e/Sr, and indications for the importance of electron-hole
Coulomb effects\cite{Ott}.

The unusual aspect of these La-doped divalent hexaborides is that for
certain values of $x$ in Ca$_{1-x}$La$_{x}$B$_{6}$ a weak ferromagnetic
moment, peaking at 0.1$\mu _{B}$/La at $x$=0.005 is found, and nearly
identical magnetic effects with La-doping of CaB$_{6}$, SrB$_{6}$ and 
BaB$_{6}$ also are found. In addition, it was recently found that Ca-deficient 
Ca$_{1-\delta}$B$_{6}$ is ferromagnetic, but the moment is distinctly smaller
than that found at $x$ = 0.005, usually by at least an order of magnitude%
\cite{Petrovic}. In all cases the maximum moment is found at x = 0.005. For
the largest moment at $x$ = 0.005 the data show the loss of magnetization
occurs with a Curie temperature near T$_{C}$ = 600 K.

Because no obvious source for a strong coupling between any possible
magnetic impurities giving rise to a Curie temperature as high as T$_{c}$ =
600 K presents itself, the coupling of defect d-state moments on this scale
seems rather implausible. One candidate for the origin of the magnetic
polarization emerges from studies of electronic correlations in the low
density electron gas, such as those of Ceperley and Alder\cite{Ceperley}.
This is a topic of theoretical speculation with a long and interesting
history, going back to Bloch and Wigner\cite{Wigner}. The study of Ceperley
and Alder is a T = 0 K computation, comparing unpolarized and completely
polarized states of the electron gas, with ferromagnetism showing up for
values of r$_{S}$, the radius of the sphere containing one conduction
electron, of order 80 a$_{B}$. For $x$ = 0.005, we compute r$_{S}$ = 15.0
\AA\ = 28.4 a$_{B}$, using the Bohr radius, a$_{B}$ for the free electron.
How the ferromagnetic ground state might occur in an ordered lattice is
unknown. The natural energy scale in this particular analysis is the Fermi
energy, and for free electrons one calculates EF = 0.062 eV = 720 K for $x$
= 0.005 in CaB$_{6}$. This temperature is of order the observed Curie
temperature. More recently, this finding has caused several theoretical
investigations into possible mechanisms for this new type of magnetism\cite
{Zhitomirsky}. In addition there have been two other recent energy band
calculations devoted entirely to this subject\cite{Rodriguez,Tromp}.

The ordered moment is approximately 0.07$\mu _{B}$/carrier at $x$ = 0.005,
which corresponds to only a partial polarization of the Fermi sea. The
reduction of the moment per carrier seen as $x$ increases and r$_{S}$
decreases might then be interpreted as a reduction in the net polarization
as the electron density increases.

We report here the results of de Haas - van Alphen measurements on 
Ca$_{1-x}$La$_{x}$B$_{6}$ and SrB$_{6}$ at mK temperatures and in fields up to 32 T in
which we observe quantum oscillations corresponding to several extremal
cross-sectional areas of the FS, proving the existence of conduction
electrons in these materials. In addition, we have measured the effective
mass of the carriers on these orbits.


These measurements were performed at the National High Magnetic Field
Laboratory, Tallahassee, FL using a torque cantilever magnetometer designed
for operation at low temperatures between 20 and 500 mK in applied fields
ranging from 0 to 32 T. Samples of CaB$_{6}$, Ca$_{0.9975}$La$_{0.0025}$B$%
_{6}$, Ca$_{0.995}$La$_{0.005}$B$_{6}$, and SrB$_{6}$ were grown in an Al
flux and etched in a 25\% HNO$_{2}$ in H$_{2}$O solution down to a small
plate that was mounted on the cantilever with GE Varnish. No evidence of Al
inclusion was observed upon magnification, and no evidence of
superconductivity from Al was observed upon cooling. The oscillatory part of
the sample's magnetization is measured as a function of field and the
resulting signal is periodic in 1/$B$ with a frequency $F$. This oscillatory
magnetization $\widetilde{M}$ is given by the Lifshitz-Kosevich (LK)
equation:

\begin{equation}
\widetilde{M}=R_{S}R_{D}R_{T}\sin \left[ \left( \frac{2\pi pF}{B} \right) -%
\frac{1}{2}\pm \frac{\pi }{4}\right] ,  \label{eq:LK}
\end{equation}

\noindent where $R_{S},R_{D}$, and $R_{T}$ are the spin, Dingle, and
temperature signal amplitude reduction factors respectively, and $p$ is the
harmonic number. The frequencies of the dHvA oscillations are proportional
to extremal areas of the FS perpendicular to the applied field direction.
The effective mass of the carriers is determined from the temperature
reduction factor by measuring the oscillatory signal amplitudes as a
function of temperature. The measured signal from the torque cantilever is a
voltage proportional to the gap between the flexible cantilever plate to
which the sample is glued and a fixed (reference) plate. The gap is measured
as a capacitance with a precision capacitance bridge. The measured
oscillations in the torque, $\widetilde{\tau }$, arise from anisotropy in
the Fermi surface, such that

\begin{equation}
\widetilde{\tau }=\frac{-1}{F}\frac{dF}{d\theta }\widetilde{M}BV
\label{torque}
\end{equation}

\noindent where $F$ is the dHvA frequency, $\theta$ is the angle of the
applied field, $B$, $\widetilde{M}$ is the LK expression above, and $V$ is
the volume of the sample. The $\frac{dF}{d\theta}$ dependence means that a
roughly spherical FS with nearly constant FS areas as a function of angle
will have a reduced torque signal from that of a highly elliptical FS.
Because the signals are inversely proportional to the measured dHvA
frequencies, this technique is particularly sensitive to the low frequencies
seen in the present results.

\begin{center}
\begin{figure}[tbp]
\epsfig{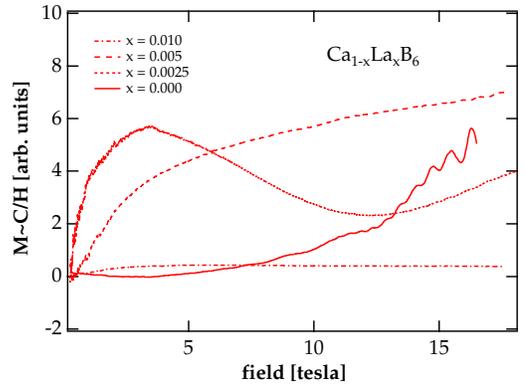}
\caption{Comparison of the relative magnetization.  The upturn in the 
magnetization of the pure \cabsix\ is due to the dHvA effect.  The $x$ = 0.0025 
trace is modulated by a low dHvA frequency.  The $x$ = 0.005 sample 
shows the largest moment.}
\label{magcomp}
\end{figure}
\end{center}

The dHvA results shown in the figures are derived from high field
measurements. The La-doped samples (except for $x$ = 0.010) showed clear
evidence of ferromagnetism as shown in Figure \ref{magcomp}. Comparing the
magnitude of the magnetization as a function of La-doping is complicated by
the low frequency dHvA, clearly seen in the pure CaB$_{6}$\ and the $x$ =
0.0025 traces. The ferromagnetic moment is maximum for $x$ = 0.005 and is
smallest for the $x$ = 0.010.

The ferromagnetic background of the $x$ = 0.0025 and 0.005 samples of 
CaB$_{6}$\ was subtracted before Fourier transforming the data. We also looked
for dHvA oscillations in an $x$ = 0.010 sample; however, none were observed.
The difficulty of observing dHvA in the $x$ = 0.010 sample can be attributed
to increased scattering.

\begin{center}
\begin{figure}[tbp]
\epsfig{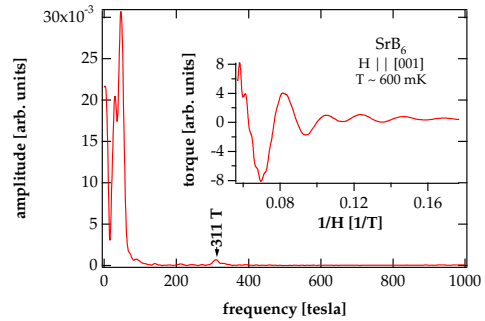}
\caption{FFT of the dHvA oscillations in \srbsix.  Inset: torque as a
function of inverse field at 600 mK.}
\label{srb6}
\end{figure}
\end{center}

Our measurements for SrB$_{6}$~ are in reasonable agreement with the
theoretically predicted FS \cite{Rodriguez} and ARPES measurements 
\cite{Denlinger}. The FS of SrB$_{6}$\ is centered around the X point and
consists of two pieces, an electron ``ring'' and a hole ``lens'' (however,
the hole pocket due to the boron bands was not observed in the photoemission
measurements by Denlinger {\em et al.}). Figure \ref{srb6} shows the raw
data and frequencies for SrB$_{6}$. The hole ``lens'' corresponds to the
dominant low frequency orbit (F $\approx$ 44 T). The smaller peak
corresponds to a FS sheet which has a cross-sectional area approximately 7
times larger than the ``lens''. This is even larger than a magnetic
breakdown orbit encompassing both the ``ring'' and the ``lens''. Neither of
the observed orbits correspond to the electron ``ring'' mentioned by
Rodriguez {\em et al.} \cite{Rodriguez}.

\begin{center}
\begin{figure}[tbp]
\epsfig{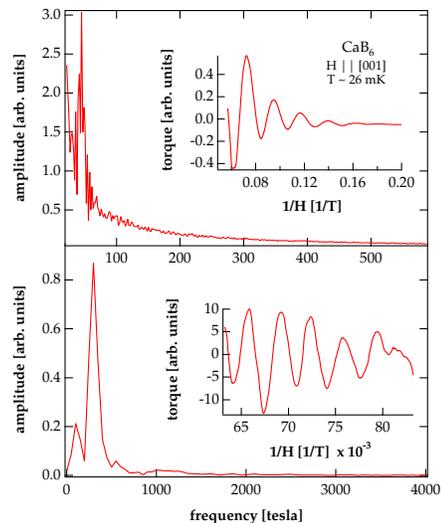}
\caption{Top: FFT and data showing the 50 T orbit \cabsix.  Bottom:
FFT and data showing the 300 T orbit.}
\label{cab6}
\end{figure}
\end{center}

The FS of pure CaB$_{6}$~ is topologically similar as can be seen in Figure 
\ref{cab6}. The higher frequency was determined by fitting the low frequency
and subtracting it from the measured data. The resulting data set showed
clear evidence of this 300 T orbit.

It should be noted that the number of holes in either CaB$_{6}$\ or SrB$_{6}$%
is determined by the number of Ca or Sr vacancies in the sample. Thus the
dHvA frequencies in these two cases may be sample dependent.

\begin{center}
\begin{figure}[tbp]
\epsfig{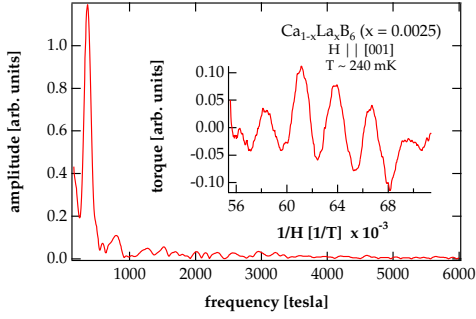}
\caption{FFT and data showing the 350 T orbit with modulation.}
\label{cala025b6}
\end{figure}
\end{center}

\begin{center}
\begin{figure}[tbp]
\epsfig{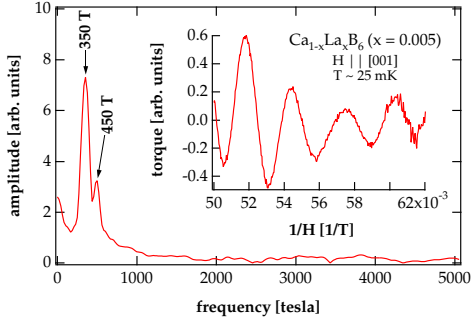}
\caption{FFT and data showing the 350 and 450 T orbits.}
\label{cala050b6}
\end{figure}
\end{center}

La doped CaB$_{6}$~ lacks the strong low frequency orbit attributed to the
hole ``lens'' in the divalent metal deficient materials; however, the higher
frequency orbit persists (see Figures \ref{cala025b6} and \ref{cala050b6}).
This higher frequency may be modulated by a frequency too low to detect. If
this is the case, then La doped CaB$_{6}$~ might have an extremely small
hole pocket. In the Ca$_{0.995}$La$_{0.005}$B$_{6}$~ sample, a 450 T orbit
accompanies the dominant 350 T orbit (Figure \ref{cala050b6}). As electrons
are added to the material, the Fermi level rises leading to a change in the
topology of the FS. This is consistent with the hole surface dropping out
with increased doping yielding a single ellipsoidal electron sheet. We
propose that the 450 T orbit arises from the ellipticity of the FS. The
eccentricity of this FS sheet is $\eta =450/350=1.3$. From the volume of the
FS the number of electrons per unit cell can be calculated \cite{EuB6}.
Because the FS is centered at the X point, 3 full FS volumes with 2 spin
directions contribute to the electron density. So assuming two states at the
FS, the electron denisty is 0.01 per unit cell. If the FS is completely
polarized with only one spin state contributing, then the density is 0.005
per unit cell.

The frequencies and cyclotron masses for each of these materials is
summarized in Table \ref{tbl:hexfreqmass}. The remarkable result showed in
this table is the mass of the higher frequencies orbits in the CaB$_{6}$ and
its electron doped variants. The cyclotron mass for the charge carriers in
SrB$_{6}$\ could not be determined because there was no measurable change in
signal amplitude from 25 to 600 mK indicating a lighter mass; higher
temperature measurements are required to determine these lighter masses. The
larger masses of the calcium materials comes from the strong
electron-electron interactions.

Because these samples were grown in an Al flux, we were careful to look for
evidence of Al contamination that might produce ``false'' quantum
oscillations \cite{Terashima}. As mentioned above, we selected small samples
which were visibly free of Al pockets. Also, we looked for the
superconducting phase transition of Al on the cool down. As a final test we
looked for dHvA signals in a piece of the Al flux on the cantilever, but the
flux showed no quantum oscillations. Additionally, the large cyclotron
masses of the higher frequency orbits mitigates against attributing their
origin to Al contamination -- there are no orbits in Al as heavy as 
4.5 m$_{e}$.

From the results of these measurements it can be seen that in every case of
HTFLCC Fermi surfaces of the conduction electrons exist. For the optimally
doped Ca$_{1-x}$La$_{x}$B$_{6}$ a single electron FS is observed and in the
divalent CaB$_{6}$ and SrB$_{6}$ an additional hole pocket is observed.
Strong electron-electron interactions are observed in each case with the
largest values observed in the optimally doped samples. Thus not only must
the mechanism giving rise to the band structure be understood, but these
inter-electron interactions must be accounted for to understand the true
mechanism giving rise to the ferromagnetism.

We would like to thank J. L. Sarrao for helpful discussions. This work was
supported in part by the National Science Foundation under Grant No.
DMR-9971348 (Z. F.). A portion of this work was performed at the National
High Magnetic Field Laboratory, which is supported by NSF Cooperative
Agreement No. DMR-9527035 and by the State of Florida.

\begin{table}[tbp]
\caption{Listing of dHvA frequencies and cyclotron masses for the
hexaborides studied in this paper.}
\label{tbl:hexfreqmass}\centering
\begin{tabular}{ccc}
{\bf Material} & {\bf F (T) $\pm$ 10} & {\bf m$^{*}$} \\ \hline
& 44 & -- \\ 
\raisebox{1.5ex}[0pt]{\srbsix} & 311 & -- \\ \hline
& 50 & 0.63 $\pm$ 0.05 \\ 
\raisebox{1.5ex}[0pt]{\cabsix} & 300 & 1.14 $\pm$ 0.18 \\ \hline
Ca$_{0.9975}$La$_{0.0025}$B$_{6}$ & 350 & 4.6 $\pm$ 0.2 \\ \hline
& 350 & 4.5 $\pm$ 0.6 \\ 
\raisebox{1.5ex}[0pt]{\calahalf} & 450 & 4.4 $\pm$ 0.9
\end{tabular}
\end{table}


\end{document}